\documentclass[10pt,journal,compsoc]{IEEEtran}

\ifCLASSOPTIONcompsoc
  \usepackage[nocompress]{cite}
\else
  \usepackage{cite}
\fi

\ifCLASSINFOpdf
  \usepackage[pdftex]{graphicx}
\else
  \usepackage[dvips]{graphicx}
\fi

\usepackage{tabularx}
\usepackage{threeparttable}
\usepackage{booktabs}
\usepackage{float}
\usepackage{xcolor}
\usepackage{authblk}
\usepackage{varwidth}
\usepackage{textcomp}

\usepackage{enumitem}
\usepackage{listings}
\usepackage{color}
\definecolor{mygreen}{rgb}{0,0.6,0}
\usepackage{hyperref}

\begin{document}

\title{CHIPKIT: An agile, reusable open-source framework for rapid test chip development}

\author[1,2]{Paul N. Whatmough} 
\author[1]{Marco Donato} 
\author[1]{Glenn G. Ko} 
\author[3]{Sae Kyu Lee}
\author[1]{David Brooks}
\author[1]{Gu-Yeon Wei} 
\affil[1]{Harvard University}
\affil[2]{Arm Research}
\affil[3]{IBM Research}


\IEEEtitleabstractindextext{%
\begin{abstract} 

The current trend for domain-specific architectures (DSAs) has led to renewed interest in research test chips to demonstrate new specialized hardware.
Tape-outs also offer huge pedagogical value garnered from real hands-on exposure to the whole system stack.
However, success with tape-outs requires hard-earned experience, and the design process is time consuming and fraught with challenges.
Therefore, custom chips have remained the preserve of a small number of research groups, typically focused on circuit design research.
This paper describes the CHIPKIT framework: a reusable SoC subsystem which provides basic IO, an on-chip programmable host, off-chip hosting, memory and peripherals.
This subsystem can be readily extended with new IP blocks to generate custom test chips.
Central to CHIPKIT, is an agile RTL development flow, including a code generation tool called VGEN.
Finally, we discuss best practices for full-chip validation across the entire design cycle.

\end{abstract}

}

\maketitle

\IEEEdisplaynontitleabstractindextext

\IEEEpeerreviewmaketitle

\IEEEraisesectionheading{\section{Introduction}\label{sec:intro}}
\IEEEPARstart{R}{esearch} test chips are the ultimate demonstration of the true value of novel computer architecture and circuits innovation. 
In addition, taping out test chips in a research or academic setting provides huge pedagogical value, offering real insight across the whole stack.
Nonetheless, taping-out test chips remains very challenging, especially for the uninitiated. 
Custom chips are time consuming to design, fabricate and test, and are error prone -- often requiring expensive re-spins to fix problems.
In this paper, we explore two key themes of \textit{agile} and \textit{reusable} design, to help reduce the barrier to entry for chip tape-outs.
Emphasizing \textit{reuse} greatly reduces development cost and at the same time minimizes the opportunity for silicon bugs, freeing the designer to focus on differentiating features.
While \textit{agile} design seeks to follow a methodology where changes can be readily implemented late into the design cycle, without significant disruption or risk.

A number of exciting and ambitious open source hardware projects currently provide an exciting range of IP blocks to use in test chip projects.
In addition to this, many IP companies are offering broad access to their products for non-commercial use.
However, turning a few IP blocks into a functioning chip that can be easily measured and debugged is still very challenging due to a significant remaining experience gap in terms of both methodology and IP blocks.

\begin{figure}[t]
        \includegraphics[width=1.0\columnwidth]{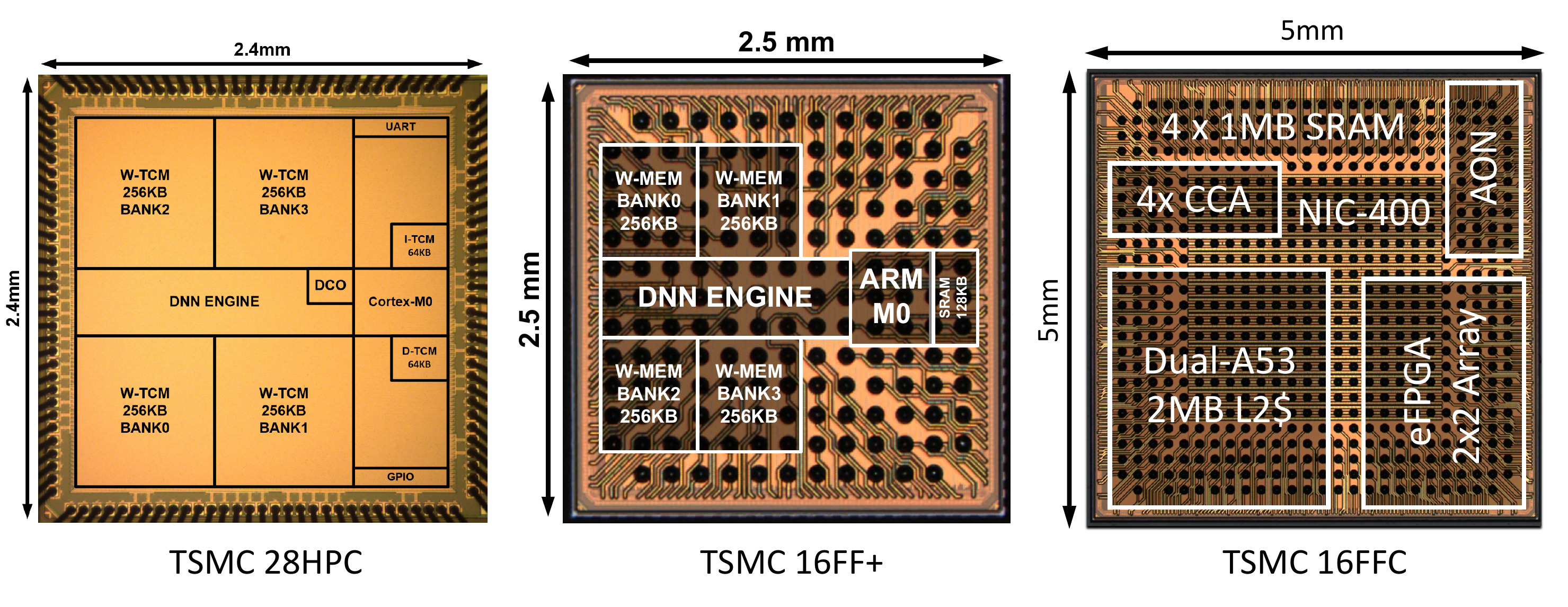}
        \vspace{-18pt}
        \caption{Three recent chips \cite{sm2-jssc,sm3-jssc,smiv-vlsi} built using the CHPIKIT framework.
        } 
        \label{fig:chips}
\end{figure}

This paper describes CHIPKIT\footnote{Available online: \url{https://github.com/whatmough/CHIPKIT}},
a straightforward framework for the rapid development of successful research test chips. 
We describe a front-to-back design example, drawing on multiple generations of successful test chips designed at Harvard (Fig.~\ref{fig:chips}), which follow a consistent design approach.
These span a range of complexities, from smaller single-accelerator micro-controller based SoCs~\cite{sm2-jssc,sm3-jssc}, through to large multi-accelerator SoCs with Arm Cortex-A multi-core CPU clusters~\cite{smiv-vlsi,sm5-vlsi}.
However, they all share the same basic framework, with the same SoC subsystem for system bring up, communication and control.
Following this framework has allowed new tape-outs to be developed with very low-risk and high success rate.
To help researchers bootstrap their own chip designs, the contents of this paper is supported with the release of our open source CHIPKIT project. 


This paper provides the following contributions:
\begin{itemize}
    \item \textbf{Reusable SoC Subsystem}, a simple on-chip CPU host, a flexible interconnect, memory, basic peripherals, and robust off-chip communication and hosting (Section~\ref{sec:soc}).
    \item \textbf{Agile RTL Development Methodology}, suitable for inexperienced designers, with robust RTL coding guidelines and a code generation tool, \textit{VGEN} (Section~\ref{sec:agile}).
    \item \textbf{Full-Chip Validation Methodology}, covering the entire design flow, which is critical to ensure functional correctness (Section~\ref{sec:validation}).
\end{itemize}


\newpage

\section{Reusable SoC Subsystem}\label{sec:soc}

\begin{figure}[t]
        \includegraphics[width=0.95\columnwidth]{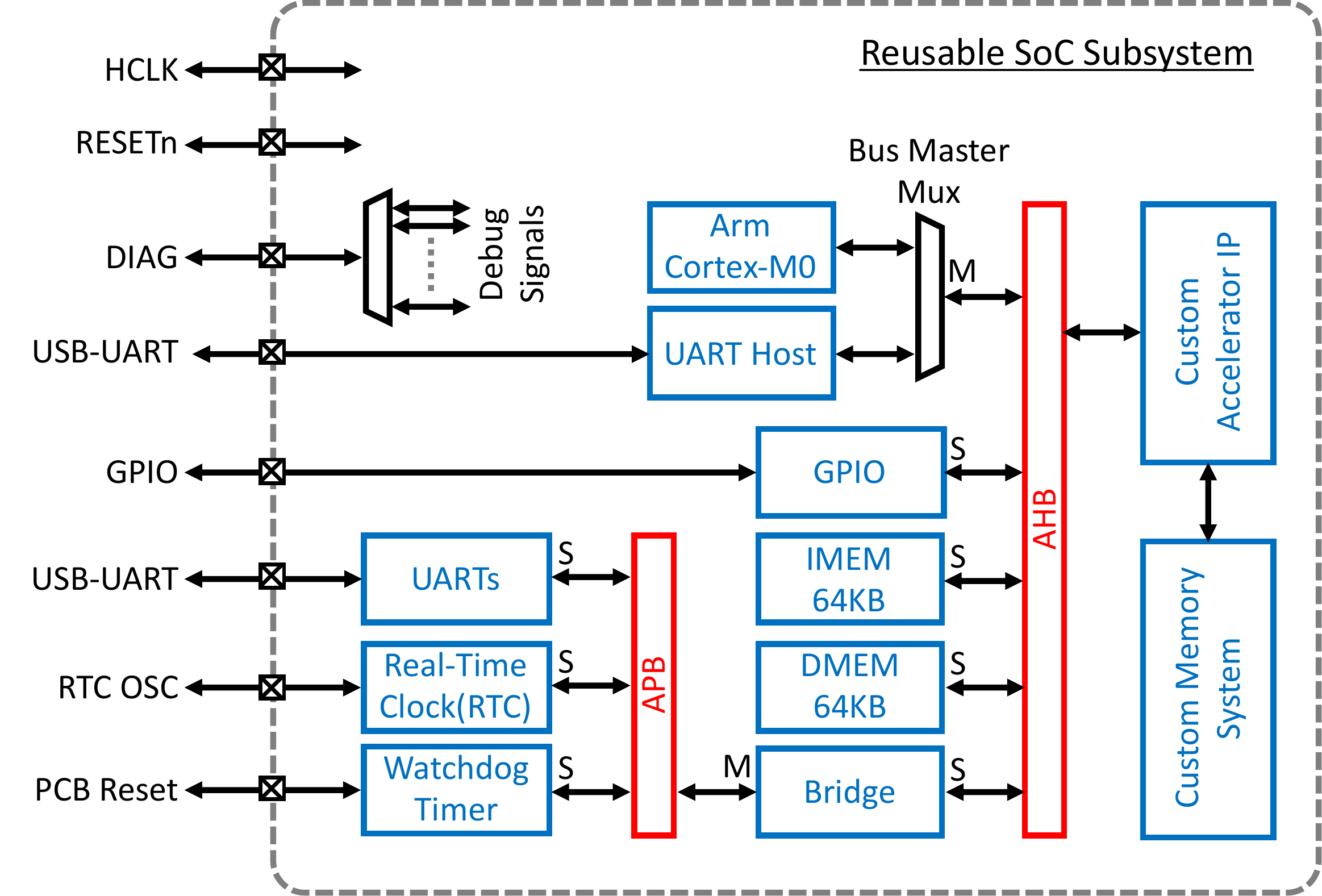}
        \caption{Reusable SoC subsystem, extended with a custom IP block.} 
        \label{fig:soc}
\end{figure}

The design goal for the reusable SoC subsystem (Fig.~\ref{fig:soc}) is to provide the minimum components to robustly handle essential bring up and test of custom IP projects.
The following subsections briefly introduce the key components.

\subsection{CPU Host}
The traditional digital chip testing approach of using external pattern generators and logic analyzers to drive and read chip pins is slow and error prone.
Instead, our subsystem includes two bus masters that can be used to run tests: a CPU and a UART master.
This configuration allows the chip to be hosted either autonomously by the on-chip CPU, or from an external PC over USB-UART.
We use a 32-bit Arm Cortex-M0\footnote{Available online: \url{www.arm.com/resources/designstart}} microcontroller, which is an extremely area efficient and easy to use design with broad software compatibility.

\subsection{UART Host}
We have developed and extensively used a simple and robust on-chip UART bus master peripheral (included in CHIPKIT) to allow an external PC host to drive transactions onto the on-chip interconnect.
This is a very useful capability for running tests, loading binaries, moving data etc.
However, it also allows whole tests to be developed and run from an external PC in any language (e.g. Python), which is a very convenient approach to chip testing.
The peripheral provides a simple interactive text interface in any standard terminal emulator, with no CPU overhead.
Single read and write transactions on-chip are initiated using simple commands, such as \texttt{R 0x70000000} to read a 32-bit word at the specified hex address.
More extensive tests are easily scripted in any language using a standard serial port library.
UART-to-USB transceivers on the PCB enable a PC to easily connect to the test chip over a USB cable.

\subsection{Interconnect}
An on-chip interconnect allows components on the SoC to communicate.
We adhere to well-documented, open industry standard bus protocols, which allows access to a broad IP ecosystem, including verification components such as protocol checkers.
In particular, we have extensively used three protocols from the AMBA standards~\footnote{Specifications available online: \url{https://developer.arm.com/architectures/system-architectures/amba}}, selected based on required features and performance: APB for low-performance peripherals, AHB where-ever possible for general-purpose, and AXI for high-performance and more features.

We typically need multiple buses, and often partition them (even on simple chips) based on usage and traffic types and volumes, which helps with throughput, as well as design and verification.
A silicon bug in an interconnect could easily hang the whole chip and therefore the interconnect IP must be robust and carefully verified.
It should also be flexible and easy to modify as the tape-out project evolves.
Open-source solutions for interfacing the components in a complex heterogeneous systems have 
been proposed~\cite{esp}. However, these solutions may go beyond the requirements of smaller SoCs used for prototyping novel DSA hardware blocks.
To better serve this purpose, CHIPKIT includes a simple single-layer AHB interconnect, which is very easy to setup and maintain.
It makes use of the SystemVerilog (SV) \texttt{interface} feature to drastically simplify RTL.
An SV interface is used to bundle the signals in each port, which can be either of type master or slave.
These bundles can then be connected using a single \texttt{modport} instance.
The address decoder in the interconnect is defined by a single SV header file which describes the entire memory map for the interconnect segment.
Adding or removing an IP is as simple as modifying the header file for the interconnect, which makes the SoC extremely agile.
An automatic default slave in the decoder catches and returns an error response for accesses to unused regions to prevent deadlocks.

\subsection{Off-Chip Interfaces}

Robust off-chip IO for control and data movement is essential for painless test chip bring up and test.
Where possible, reusing the same basic off-chip signal IO arrangement reduces risk and development effort.
The essentials include an off-chip clock and reset, UARTs, general purpose IO (GPIO), real-time clock (RTC) oscillator, diganostic (DIAG) signals, and any CPU debug interfaces.

When using the on-chip CPU host, \texttt{printf()} calls can be retargeted to the UART slaves.
This is also useful in RTL simulation, where the CPU can terminate a test at the end of a program by writing a unique ASCII code to the slave UART which is used by the test bench to end simulation.

%

It is often necessary to debug issues during test chip bring-up, which can be very challenging in silicon due to lack of visibility.
To help increase visibility, we include a diagnostic (DIAG) pin multiplexer to allow multiple signals to be observed off-chip (Fig.~\ref{fig:soc}), without requiring a large number of chip pins.
The mux select signal is controlled by a memory-mapped register.
Typical signals to observe include clocks, resets, power rails, power gate enables, FSM states, interrupts etc.
Using at least two DIAG pins allows relative observation, e.g. observe clock and reset at the same time.
The DIAG multiplexers themselves can be trivially implemented and automatically populated with signals using a VGEN script (Section~\ref{sec:agile}).

\subsection{On-Chip Memories}

On-chip SRAM memories are included for storing binary programs and data.
ROMs may also be useful, such as for a boot loader.
To be bus accessible, SRAM and ROM macros require a bus interface, and CHIPKIT includes a suitable AHB SRAM interface.



Control and status registers (CSRs) are very common at both the SoC and IP level.
In particular, research test chips tend to include a larger number of CSRs in order to configure experiments and turn them on or off (so-called chicken bits).
CSRs can quickly become very time-consuming to implement, document, modify and maintain.
CHIPKIT provides an agile flow to automatically generate CSRs from a single database using a VGEN Python script, which we will describe in Section~\ref{sec:agile}.
This approach makes it very convenient to add CSRs as needed during RTL design.


\subsection{Peripherals}
The SoC includes some basic low-bandwidth peripherals arranged on a compact APB interconnect.
These include UART slaves, a watchdog timer, and a real-time counter (RTC).
The RTC is especially useful for measuring the runtime of a given workload while characterizing the chip.

\subsection{Clocks, Resets and Power}

A simple and robust clock and reset architecture is essential in research test chips to prevent potential complications in what is a truely essential aspect of digital electronics.
A single off-chip clock (\texttt{HCLK}) and asynchronous active-low reset (\texttt{RESETn}) is supplied to the chip from the PCB (Fig~\ref{fig:soc}).
Due to the controlled slew rate of standard IO cells, \texttt{HCLK} is typically limited to a maximum of a few hundred MHz.
Any faster clocks must be generated on-chip.
A straightforward approach to generating fast clocks on-chip is to used an open-loop digitally-controlled oscillator (DCO), which can be implemented using a netlist of digital standard cells, without any custom hand layout.
Multiple power domains are useful on test chips, to measure power consumption of individual blocks, or to perform fine-grained dynamic voltage and frequency scaling (DVFS).
However, power domains add a huge amount of complexity in both RTL development and (especially) implementation, which brings risk.
For research test chips, we suggest using a lightweight approach; in particular it is a good idea to avoid the use of power-gates and level-shifters, which add a huge amount of complexity in the EDA flow, along with validation overhead.
This approach is usually feasible if the voltage ranges are sufficiently close and there is no strong requirement to power-off domains.

\subsection{Adding Custom IP Blocks}
The SoC subsystem is a foundation upon which research test chips can be rapidly constructed by adding new IP blocks~\cite{sm2-jssc,sm3-jssc} or even whole subsystems~\cite{smiv-vlsi}.

The method of interfacing new blocks will largely depend on the complexity of the IP to be added.
The simplest approach is a slave programming model, where a software driver programs the accelerator with data and control information, before initiating the accelerator, which executes the task and returns an interrupt on completion.
Higher-performance blocks will require a more sophisticated interface, such as including a master bus interface on the accelerator to allow it to initiate data transfers independently.
In fact, programming models and interfacing of accelerators is a very active area of research, especially considering things like data movement cost, virtualization and coherency~\cite{acai,smaug}.




A fast clock domain will allow an accelerator to achieve higher performance.
Note that this introduces an asynchronous or isochronous clock-domain crossing (CDC) around the bus interface, which will require a CDC bridge to ensure correct data transfer.

%

\section{Agile RTL Development}\label{sec:agile}

Research test chip projects are typically severely time constrained.
Therefore, it is important to use an RTL development approach that is 1) efficient, 2) minimizes opportunity for bugs, and 3) is supported by front-to-back EDA tool flows.
In recent years, there has been a significant research effort exploring new hardware design languages~\cite{chisel_dac12,pymtl_micro14}, as well as high-level synthesis (HLS) from C++/SystemC~\cite{nvidia_hls}.
Chipyard\footnote{Available online: \url{https://github.com/ucb-bar/chipyard}} provides an comprehensive SoC design framework in Chisel.
In contrast, CHIPKIT focuses on parameterized SystemVerilog (SV) for RTL design.
Compared to Chisel, SV is mature, natively supported by EDA tools~\cite{sv_snug13}, and relatively well supported~\cite{basejump}.
SV also does not require an opaque translation step to generate RTL for simulation and implementation.
Which can speed up validation and implementation, which tend to far exceed the time originally spent on design.

We have found that the quality of RTL coding in academia is often poor, especially in comparison to industry RTL.
Poor RTL can lead to long debug cycles and is time conusming to maintain and update.  It can even lead to silicon bugs.
We have found that introducing strict coding guidelines can effectively solve this problem.
Therefore, in this section, we discuss the components of an agile RTL development process, that uses standard commercial simulation and implementation tools.

\subsection{SystemVerilog Coding Style}
\label{sec:agile:coding}

\begin{figure}
\centering

\begin{lstlisting}
`include RTL.svh

module my_counter (
  input  logic       clock,
  input  logic       reset_n,
  
  input  logic       enable,
  output logic[31:0] count
);

// Use "logic" type exclusively, not "wire" or "reg"
logic[31:0] count_next;

// Use "always_comb" keyword for logic
always_comb count_next = count + 32'd1;

// Use a macro to infer registers
`FF(count_next, count, clock, enable, reset_n, '0); 

endmodule  // my_counter
\end{lstlisting}

\vspace{-8pt}
\caption{SystemVerilog coding guidelines example.}
\label{fig:sv}
`\end{figure}

SV is a very large language with many verification-oriented features that are not relevant to writing synthesizable RTL.
Therefore, we use a strict RTL coding style, which can be summarized in the following directives:

\begin{itemize}
\item \textbf{Separate logic and registers.}
Makes RTL easier to parse and pipelining easier to modify.
\item \textbf{Use rising-edge registers with active-low async reset.}
Simplifies timing constraint development.
\item \textbf{Use the \texttt{logic} type exclusively.}
Replaces both the older \texttt{wire} and (very confusing) \texttt{reg} types.
Provides compile-time checking for multiple drivers.
\item \textbf{Use the \texttt{always\_comb} keyword for logic.}
Provides compile-time checking for unintended latches or registers.
\item \textbf{Use the \texttt{always\_ff} keyword to infer registers.}
Provides compile-time checking for unintentional latches.
\item \textbf{Use automatic module instantiation connections (\textit{.*}).}
These significantly reduce the verbosity of connecting modules and provide additional compile-time checking.
\end{itemize}

In addition to these guidelines, we also recommend the strict use of a pre-processor macro for register inference.
This has a number of advantages, including: 1) significant reduction in lines of code, 2) removes the risk of poor inference style, 3) enforces use of a rising-edge, async active-low reset, 4) allows the register inference template to be changed to suit ASIC or FPGA.
A macro is used instead of a module to reduce simulation overhead.
Fig.~\ref{fig:sv} gives an example module for a simple counter, following our guidelines, including the use of the CHIPKIT SV header (\texttt{RTL.svh}) which includes a register macro (\texttt{`FF()}).


\subsection{Instantiated Library Components}
\label{sec:agile:libs}

Physical IP such as SRAMs, IO cells, clock oscillators, and synchronizers need to be instantiated in the RTL.
It's worth remembering that various versions of these cells may be required over the lifetime of the IP or full-chip, including RTL functional models as well as various ASIC and FPGA libraries.
Therefore, it is helpful to wrap instantiated components inside a module, which can then be easily switched.
Each set of wrapped component instantiation modules is stored in a different directory for each library, with the correct directory included at compile time.

\subsection{VGEN Code Templating}
\label{sec:agile:vgen}

\begin{figure}[t]
    \centering
    \includegraphics[width=0.95\columnwidth]{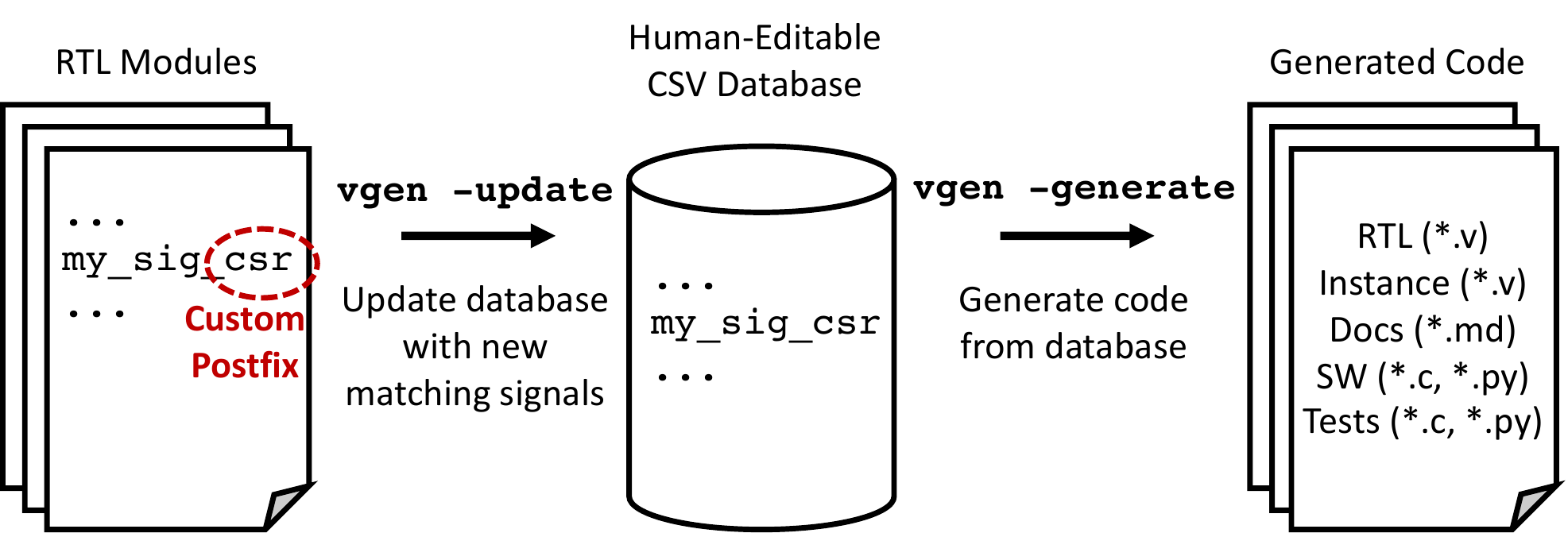}
    \caption{VGEN agile templating flow for the CSR generation example.}
    \label{fig:vgen}
\end{figure}

In chip development, it is common to encounter repetitive coding tasks.
Typical examples include SoC memory-maps, CSRs, IO signals/pads, clocks and resets and debug signals.
These can be tedius and error-prone and introduce significant risk into the tape-out project.
The perennial preferred approach is to write \textit{generators}.
CHIPKIT includes \textit{VGEN}, which is a simple Python framework for writing templated code generators.

As a prototypical example, consider the implementation of CSRs (Section~\ref{fig:soc}).
Whenever a new CSR is added during the design process, the following need to be updated: 1) RTL, 2) documentation, 3) C/Python software views, 4) CSR tests.
A typical CSR module with 100 register definitions requires over a thousand lines of RTL to be written, maintained and validated.
Adding a new CSR therefore, becomes an extremely time-consuming and error-prone process. 
VGEN automates the generation of all this code requiring only 126 lines of code.

Fig.~\ref{fig:vgen} gives an outline of the VGEN flow, which operates in two stages.
The first step is to update a CSR database with signals from the design, which can be done periodically as the RTL is developed.  
The VGEN tool automatically updates the database (\texttt{vgen -update}) by parsing RTL modules to find signal names with a matching prefix or postfix that indicates a CSR should be attached to the signal.
Any matching signals are then cross-referenced against the database to see if they already exist and if any extracted parameters, such as the bitwidth of the register have changed.
If it is a new addition or a modification, the change will be made in the database.
The database is stored in comma-separated value (CSV) format, which allows it to be easily viewed and edited in a spreadsheet program.
The CSV database can be version controlled alongside the RTL.

The second stage of operation is to proceed and generate templated output code with values from the database (\texttt{vgen -generate}).
For CSRs, an RTL module is generated with memory-mapped registers as described in the CSV database, along with code for a module instantiation template.
Documentation in Markdown format is also generated, along with C and Python software register definitions and tests to confirm correct operation of the automatically generated code.

VGEN is a lightweight Python module.
The database data structure is represented as a list of dictionaries.
The keys for the dictionaries are defined in the header line of the CSV file, so it is easy to add new attributes by simply editing the CSV file.
CHIPKIT currently includes example VGEN scripts for generating CSRs and IO pads, and is easily extensible to other common chip design tasks.


%

\section{Test Chip Validation}
\label{sec:validation}

Validation effort typically far outweighs design, and should thus be a primary consideration.
Although this is a huge topic, the following section presents some general advice and guidelines to help ensure first time right silicon for research test chips.

\subsection{IP-Level}
As a first step, documentation should be considered an essential form of IP validation.
Adopting a Markdown format for documentation allows it to be developed and version controlled alongside the RTL.
As a priority, documentation should also include a detailed block diagram of the IP.
Another useful first step is to implement an \textit{integration shell}, which has the interface signals described in the documentation.
This is very useful for preliminary SoC development.
It should compile without error, but typically does not include much, if any, functionality.
As the IP matures, the integration shell can be replaced with the full RTL, accompanied by suitable integration tests.

RTL simulation at various granularities, is the workhorse of IP development.
This typically involves a testbench, a set of self-checking tests and an associated Makefile or script to run them.
Smaller ad-hoc unit-tests and RTL assertions should also be developed as new modules are coded.

As the IP approaches the feature-complete milestone, the timing and power closure process will proceed, and a greater breadth and depth of validation will help improve the quality of the design.
Simulator coverage tools give a good indication of the maturity of test suits and which parts of the IP warrant further attention.
Linting tools provide a static check for RTL coding errors and clock-domain crossing issues.
Other static tools can help with optimizing clock gate enables and RAM enable efficiency.
Early synthesis trials will help flush out long timing paths in the design.
Power analysis will give an indication of the power consuming blocks in the design, indicating targets for further optimization.

%

\subsection{SoC-Level}
At the SoC level, many of the same guidelines discussed for the IP level apply.
However, there are also a greater number of details that tend to make it difficult to achieve high test coverage.
Top of the list for validation scrutiny is everything required for initial bring up, including clocks, resets, power sequencing (if any), and basic off-chip communication.
It is essential that these design components work without fail, and the boot sequence must be carefully validated before tape-out.
The interconnect is another critical area for validation.
A comprehensive, automatically-generated test is useful to check correct operation of all regions in the memory map, both valid and unmapped.
For large memories, rather than just testing the first few words, be sure to toggle all address and data bits to catch accidental signal truncation in RTL hierarchies, which is common.
At the SoC-level, there is a trade-off between coverage and run-time, so it may be necessary to optimize big tests to achieve the best coverage in reasonable run time.
Finally, it's important to re-run IP-level tests on the SoC to check correct interface and functionality assumptions hold.
The ultimate goal is to run all the tests in simulation that you intend to measure on the silicon.

With some additional effort, the accuracy with which RTL simulation models real circuits can be enhanced.
A good example of this is to setup an option to run simulation regressions with undefined SRAM initial states (``X'' values), which removes dependence on SRAM power-up states, which will be random on real silicon.
Similarly, clock domain crossings (CDCs) are a big concern in this regard.  Hence, another useful simulation capability is to set any asynchronous clocks to be randomly jittered in period relative to each other, which helps to catch CDC paths with missing synchronizers.
If possible, reset synchronizers should be avoided in favor of individually controllable resets, which provide greater control for debug.

\subsection{Design for Test and Debug}
The test and (occasional) debug process will be much smoother if carefully considered at design time.
The IP should include instrumentation to perform any measurements required to demonstrate and measure the expected operation.
This typically means some kind of performance counters, usually manipulated using CSRs.
For example, clock cycle or memory access counters provide essential data to characterize performance.
Some kind of infinite loop or autonomous self-test mode can also be useful to allow easy average power measurements without including any data loading phase that is otherwise required to test the design.

It's almost inevitable that at some point it will be necessary to debug unexpected behaviour on real silicon.
Debugging any design aspect that includes significant complexity can be very challenging partly due to limited visibility in real silicon. 
Therefore, this should be considered during design time.
The DIAG mux approach described in Section~\ref{sec:soc} is a cheap way to provide visibility from outside the chip, and should include all clocks, resets and other critical hardware state.
Full control of clocks and resets from the SoC by means of dedicated CSRs is also essential.
Finally, it is also a good general rule to make all SRAMs and register files in the chip memory mapped.
Although this adds additional complexity, it helps when writing self-checking tests and is invaluable when debugging SRAM circuit performance.

\subsection{FPGA Emulation}
Successfully validating a test chip RTL codebase on an FPGA will drastically increase the chances of success on ASIC.
The process of running the RTL through an FPGA toolflow can uncover a myriad of functional and timing issues.
As well as helping uncover bugs in the RTL, FPGA implementation will also help with timing constraint debug.
The FPGA emulation will also be significantly faster than RTL simulation, which enables much more extensive stress test regressions.
Finally, FPGA emulation is a great chance to check the correct operation of off-chip interfaces with the opportunity to run a more convincing end-to-end test, without any ``magic'' help from testbenches which can do things in simulation that are not possible on a real chip, such as pre-loading SRAM in zero time.
The RTL coding guidelines presented in Section~\ref{sec:agile:coding} along with the guidelines for instantiated library components in Section~\ref{sec:agile:libs} should make porting the SoC to FPGA straightforward.

\subsection{Physical Design}
Physical design is the final process before tape-out, and is obviously a critical focus for full-chip validation, which we will discuss briefly here from the RTL design perspective.

As soon as the RTL codebase will compile, the physical design flow can be developed, starting with developing timing constraints in synthesis.
Beyond this, static timing analysis (STA) reports from the implementation flow provide the feedback for RTL timing closure.
This process is iterative, as refactoring logic and pipelines to reduce the gate delays on a path tends to reveal other near-critical paths which need attention.
Therefore, it is helpful to start using STA early in the design process as the microarchitecture solidifies.
Throughout this process, VGEN code generation (Section~\ref{sec:agile:vgen}) can be used to automatically generate repetitive design-specific scripting, such as IO pad placement scripts, as the project evolves.

As the design matures, the validation effort in the back-end will ramp.
Logical equivalence check (LEC) tools allow synthesis and layout netlists to be formally checked against the RTL.
These netlists should also be simulated in the RTL testbench, using the library vendor Verilog models of cells and SRAMs.
Netlists are simulated in various degrees, including without annotating delays to cells and wires (\textit{zero-delay} mode), through to full annotation with dynamic timing checks.
The former (zero-delay) is relatively easy to setup and should be included in regressions early in the validation stage.
The latter (full timing annotation) can take some work to get running.
However, annotated netlist simulation (over PVT corners) is essential to help debug any potential errors in the timing constraints, which will not be caught by STA alone.

\section{Conclusion}
Chip tape-outs can be time consuming and error prone.
This paper describes \textit{CHIPKIT}, an agile reusable framework for rapidly developing robust test chips.
The basis of this framework is a straightforward SoC subsystem that provides basic IO, communication and both on and off -chip hosting.
Research chips with various experiments can readily be built on top of this, without having to reinvent the wheel each time.
With the current interest around domain-specific accelerators, we believe the CHIPKIT framework will enable more research teams to demonstrate their work in custom silicon.

\section{Acknowledgements}
This work was supported by the Applications Driving Architectures (ADA) Research Center, a JUMP Center co-sponsored by SRC and DARPA\@.
\ifCLASSOPTIONcaptionsoff
  \newpage
\fi

\bibliography{main.bib}{}
\bibliographystyle{ieeetr}

\vskip -2\baselineskip plus -1fil
\begin{IEEEbiographynophoto}{Paul N. Whatmough}
leads research on hardware for machine learning at Arm Research Boston, and is an Associate at Harvard University.
His research interests include efficient algorithms, computer architecture, and circuits.
Whatmough received a PhD in electrical engineering from University College London, U.K.
Contact him at paul.whatmough@arm.com.
\end{IEEEbiographynophoto}

\vskip -2\baselineskip plus -1fil
\begin{IEEEbiographynophoto}{Marco Donato}
is a Research Associate at Harvard University. His research interests include novel design methodologies targeting energy-efficient, reliable circuits and architectures for emerging computing paradigms. Donato received a PhD in electrical engineering from Brown University. Contact him at mdonato@seas.harvard.edu.
\end{IEEEbiographynophoto}

\vskip -2\baselineskip plus -1fil
\begin{IEEEbiographynophoto}{Glenn G. Ko}
is a postdoctoral researcher at Harvard University. 
His research interests include machine learning, algorithm-hardware co-design and scalable accelerator architectures on the cloud and edge.
Ko received a PhD in electrical and computer engineering
from the University of Illinois at Urbana-Champaign and
worked on Samsung Exynos SoCs prior to that. Contact 
him at gko@seas.harvard.edu.
\end{IEEEbiographynophoto}

\vskip -2\baselineskip plus -1fil
\begin{IEEEbiographynophoto}{Sae Kyu Lee}
is a Senior Hardware Engineer at IBM T.J Watson Research Center. 
His research interests include circuits, architecture and design methodologies for energy-efficient accelerators. 
Lee received a PhD in electrical engineering from Harvard University. Contact him at saekyu.lee@ibm.com.
\end{IEEEbiographynophoto}

\vskip -2\baselineskip plus -1fil
\begin{IEEEbiographynophoto}{David Brooks}
is the Haley Family Professor of Computer Science at Harvard University. His research interests include architectural and software approaches to address power, thermal, and reliability issues for embedded and high-performance computing systems. Brooks received a PhD in electrical engineering from Princeton University. Contact him at dbrooks@seas.harvard.edu.
\end{IEEEbiographynophoto}

\vskip -2\baselineskip plus -1fil
\begin{IEEEbiographynophoto}{Gu-Yeon Wei}
is a Gordon McKay Professor of Electrical Engineering and Computer Science at Harvard University. His research interests include mixed-signal integrated circuits, computer architecture, and runtime software, looking for cross-layer opportunities to develop energy-efficient systems. Wei received a PhD in electrical engineering from Stanford University. Contact him at guyeon@seas.harvard.edu.
\end{IEEEbiographynophoto}

\end{document}